\documentclass[a4paper]{jpconf}
\usepackage{graphicx}
\usepackage{verbatim}
\usepackage{amsmath}
\usepackage{epstopdf}

\begin{document}
\title{Chemically gated electronic structure of a superconducting doped topological insulator system}

\author{L A Wray$^{1,2}$, S Xu$^2$, M Neupane$^2$, A V Fedorov$^1$, Y S Hor$^{3,4}$, R J Cava$^4$ and M Z Hasan$^2$}
\address{$^1$Advanced Light Source, Lawrence Berkeley National Laboratory, Berkeley, California 94305, USA}
\address{$^2$Department of Physics, Joseph Henry Laboratories, Princeton University, Princeton, NJ 08544, USA}
\address{$^3$Department of Physics, Missouri University of Science and Technology, Rolla, MO 65409}
\address{$^4$Department of Chemistry, Princeton University, Princeton, NJ 08544, USA}

\begin{abstract}

Angle resolved photoemission spectroscopy is used to observe changes in the electronic structure of bulk-doped topological insulator Cu$_x$Bi$_2$Se$_3$ as additional copper atoms are deposited onto the cleaved crystal surface. Carrier density and surface-normal electrical field strength near the crystal surface are estimated to consider the effect of chemical surface gating on atypical superconducting properties associated with topological insulator order, such as the dynamics of theoretically predicted Majorana Fermion vortices.

\end{abstract}

Three-dimensional topological insulators (TIs) are a class of materials distinguished by topological parameters of bulk band structure, and by the consequent appearance of distinctive spin-helical surface states \cite{Intro,TI_RMP,TIbasic,DavidNat1,DavidScience,DavidTunable,MatthewNatPhys}. When combined with superconductivity, the surface environment of TIs is expected to host effective Majorana Fermion particles with collective non-Abelian statistics. Electron doped Bi$_2$Se$_3$ has been identified as one model system in which superconducting phase coherence can be induced in TI surface states \cite{WrayCuBiSe,WrayCuBiSePRB,AndoSC,dongBiSeFilmSC}. In this study, copper atoms are deposited on the cleaved surface of N-type Cu$_x$Bi$_2$Se$_3$ crystals to better understand how the surface energetics of Bi$_2$Se$_3$ can be modified, and to explore the material as a potential platform for the manipulation of Majorana vortex modes. Angle resolved photemission spectroscopy is used to observe surface energetics and the formation of new surface-localized electronic states. Surface-normal charge density distributions are estimated using the modified Thomas Fermi approximation (MTFA) \cite{MTFA}.

\begin{figure}[t]
\centering
\includegraphics[width = 8.5cm]{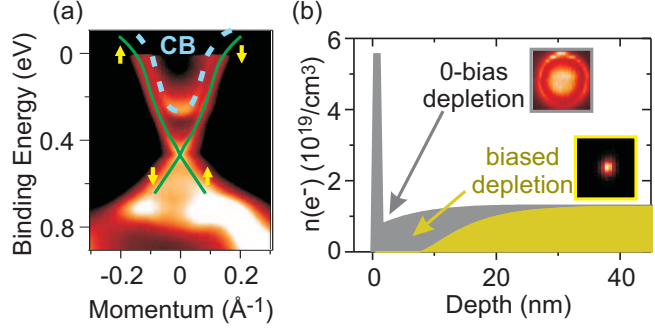}
\caption{{\bf{Intrinsic charge inhomogeneity at a doped TI surface}}: (a) (blue) Bulk and (green) surface bands intersecting the Fermi level of Cu$_x$Bi$_2$Se$_3$ with x=0.12 superconducting composition are traced above ARPES measurements. A diagram above the ARPES image shows topological quantum numbers and the location of the topological surface states. (b) The electron density distribution is modeled for (gray) as-grown Bi$_2$Se$_3$ and (gold) the same sample with chemical potential set at the Dirac point by NO$_2$ deposition. Even with no surface deposition, large electron density in the surface state (top nanometer) is expected to create a minor depletion dip in charge density. Insets show the Fermi surface, from which the surface state charge density is determined using the Luttinger theorem.}
\end{figure}

\begin{figure*}
\centering
\includegraphics[width = 11cm]{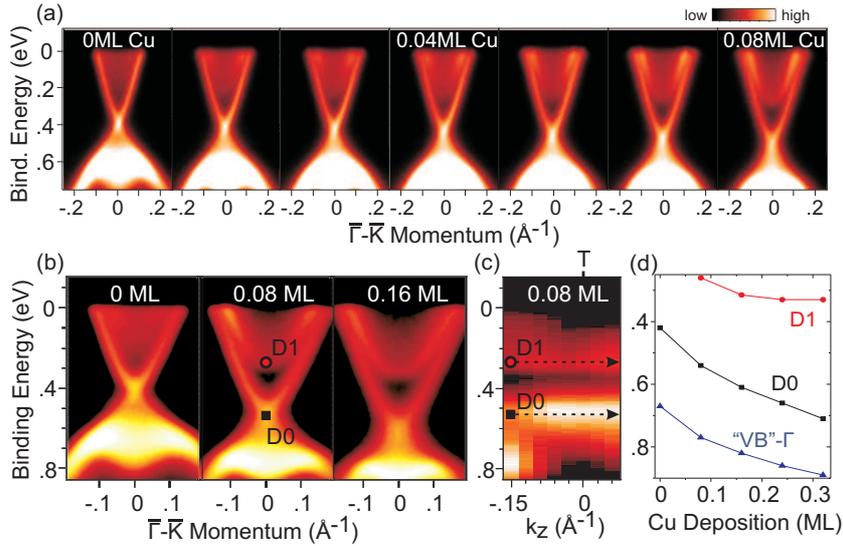}
\caption{{\bf{Cu deposition on Cu$_x$Bi$_2$Se$_3$}}: (a) Photoemission measurements show the incremental formation of a 2DEG surface state as an estimated 0.08ML of Cu is added to the surface of bulk-doped Cu$_x$Bi$_2$Se$_3$. (b) Deposition is shown over a wider range. (c) The new band that appears after 0.08 ML Cu deposition, labeled `D1' at the BZ center, has no momentum dispersion along the z-axis. The original surface state Dirac point is labeled `D0'. (d) The doping evolution of band energies at the $\Gamma$-point is shown.}
\end{figure*}

Bismuth selenide has a bulk band gap of $\Delta$$\sim$0.3eV, which is spanned by a single surface state Dirac cone at the 2D reciprocal space $\overline{\Gamma}$-point (Fig. 1(a)). Light copper doping does not significantly change the bulk band structure of Cu$_x$Bi$_2$Se$_3$ \cite{WrayCuBiSe,WrayCuBiSePRB}. A highly variable renormalization of surface electron velocities is found as a function of carrier doping, and has been suggested to be linked to the chemical potential rather than deriving from hybridization with chemical impurities \cite{WrayPN}. Electrostatic repulsion and quantum exclusion from the topological surface state shift the energies of bulk-derived electronic states near the surface \cite{fisherBending,WrayCuBiSePRB,WrayPN}, and are expected to create a region with slightly depressed carrier density even when no external bias is applied (Fig. 1(b)).

Bulk electron charge density is calculated using MTFA from the measured band dispersions and surface carrier density for pre-deposition samples, matching the approximations adopted to calculate band bending energies in Ref. \cite{fisherBending}. After deposition, surface charge is estimated from a combination of the ARPES Luttinger count and the deposited density of Cu atoms. These calculations are intended as qualitative illustrations only, and therefore make use of the nominal dielectric constant of $\epsilon_r$=113 \cite{dielConst} as opposed to attempting to treat $\epsilon_r$ more accurately as a function of depth inside the crystal. Additionally, the Dirac cone surface state of freshly cleaved Bi$_2$Se$_3$ is assumed to be localized in the top nanometer of the crystal, consistent with DFT simulations and experimentally base estimates \cite{WrayCuBiSePRB}. Similar ARPES-based methodologies for exploring the effect of electrostatic perturbations on TI surfaces have been established in previous studies \cite{DavidTunable,MatthewNatPhys,fisherBending,WrayFe,HofmannRashba2DEG,PanNN,WrayPN}.

\emph{In-situ} copper deposition and concurrent ARPES measurements were performed at Advanced Light Source beamline 12, at temperatures lower than 20K to minimize migration of Cu atoms into the sample. ARPES energy resolution was better than 20 meV, and the vacuum was maintained below 8$\times$10$^{-11}$ Torr. Deposition depth is presented in units of monolayers (ML) to approximately represent the number of deposited atoms per hexagonal unit cell of the surface, based on rough linear assumptions outlined in Ref. \cite{WrayPN}.

Intensity resembling a new surface state appears near the Fermi level after deposition of just $\sim$0.01ML Cu on the surface of N-type Cu$_{0.02}$Bi$_2$Se$_3$ (Fig. 2(a)). After 0.08ML deposition, the new surface band is clearly seen at all momenta beneath the Fermi level, and shows no z-axis momentum dispersion (Fig. 2(c)). Momentum along the z-axis is estimated as in Ref. \cite{WrayCuBiSe,WrayCuBiSePRB}. Excess surface-deposited Cu adatoms cause the electron energies near the surface of the TI to bend towards higher binding energy, as they would in an N-N Schottky diode (Fig. 2(d)).

Because TI surface states cross the bulk band gap and have different energetics than bulk electrons \cite{TIbasic,DavidScience}, when interfacing non-TI materials with a topological insulator, the surface states are expected trap charge carriers and attract or repel bulk electrons even when the bulk carrier doping of each material is identical. The surface charge density evaluated from the Luttinger count suggests that prior to depositing additional copper on the vacuum-cleaved Cu$_{0.02}$Bi$_2$Se$_3$ surface, a negative charge is trapped by the surface state, and will be screened over a range of more than 10 nanometers as simulated by MTFA (Fig. 3(a)). Approximating +1 valence for Cu at low coverage, the simulation suggests that depositing 0.08ML of copper causes the net charge at the crystal surface to undergo a significant change from negative to positive. The approximation of nearly full +1 ionization of surface Cu atoms is known to break down at higher coverage \cite{WrayPN}. Between 0 and 0.08ML (shaded in Fig. 3(a)) this picture suggests that the electric field penetrating beyond the first few nanometers of the crystal surface is very small, allowing electrons to retain their bulk-derived energetics and symmetries as they approach the surface from deeper in the crystal. The existence of a sign change in the surface charge is supported by the slower change in surface energies when adding Cu from 0.08ML to 0.16ML surface coverage, relative to the first 0.08ML of deposition. Titrating with electropositive adatoms or a gating voltage (V$_G$$\sim$0.1V from our data) to achieve an electrically neutral surface may therefore provide a method to enhance bulk interactions with the surface states, such as magnetic and superconducting proximity effects.

\begin{figure*}[t]
\centering
\includegraphics[width = 11cm]{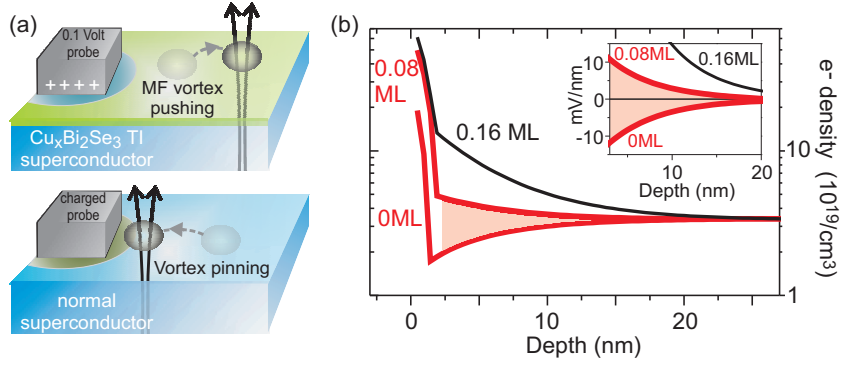}
\caption{{\bf{Majorana vortices and surface bias}}: (a) Probes interfacing with normal superconductors suppress the superfluid density, attracting (``pinning") vortices. On a topological insulator, a gentle probe with bias voltage similar to 0.1eV can potentially repel vortices by enhancing the superfluid density. Regions with suppressed superfluid density are indicated with green shading. (b) The z-axis electron density distribution is modeled for different degrees of surface doping. (shaded region) Deposition levels between 0ML and 0.08 ML are expected to reduce bulk screening and strengthen superconductivity near the surface. An inset shows the MTFA estimation of electric field strength inside the crystal.}
\end{figure*}

When superconducting phase coherence is introduced at the surface of a TI, effective Majorana Fermion modes are expected to manifest where magnetic vortex lines intersect the surface states \cite{FuSCproximity}. These Majorana states can give rise to non-Abelian statistics, and may be possible to manipulate adiabatically to store quantum information, given the right combination of superconducting order parameter and chemical potential \cite{WrayCuBiSe,WrayCuBiSePRB,HosurVort}. There is strong evidence that Cu$_x$Bi$_2$Se$_3$ (x$\geq$0.1) is a so-called topological superconductor \cite{WrayCuBiSe,WrayCuBiSePRB,AndoTopoSC,FuTopoSC2012}, which would create gapless states that preclude adiabatic manipulation of the surface. However, bismuth selenide thin films can also allow standard s-wave superconducting phase coherence to propagate into surface states from a niobium diselenide substrate \cite{dongBiSeFilmSC}, satisfying the known prerequisites for adiabatic manipulation of Majorana Fermions.

The MTFA calculation in Fig. 3(a, inset) provides an estimate that the electric field from the negatively charged surface state will have an energy scale per nanometer that is large relative to the 2$\Delta$$\sim$1meV superconducting gap function in the top $\sim$20nm of the crystal \cite{WrayCuBiSe,WrayCuBiSePRB,AndoSC,AndoTopoSC}. This is expected to cause phase fluctuations that suppress superconductivity if the surface charge is not neutralized. Moreover, because vortices are repelled by regions of high superfluid density, this surface environment provides a new mechanism to control the dynamics of Majorana fermions and potentially observe their non-Abelian properties. For normal superconductors, surface perturbations suppress superfluid density, generating an attractive force that pins vortices in place \cite{VortexPin,VortexPush}. For a topological insulator however, our data show that adjusting the surface chemical energy to cancel out the negative surface state charge can likely enhance superfluid density under a probe, causing surface vortices to be repelled by the probe rather than pinned to it (Fig. 3(b)).

It should be emphasized that the MTFA calculations we have discussed are provided only as a basis for qualitative discussion. We have not gone to great lengths to make these calculations quantitatively accurate, as that would be very difficult for such a complex surface environment. Error relating to the z-axis distribution of dielectric permeability and z-axis localization of the newly bound 2DEG-like surface states is also difficult to estimate. We have assumed that the newly bound surface state is highly localized in the top 2 nanometers of the crystal, which provides a sufficient minimum basis for describing the states in a k.p theory derived model \cite{WrayPN}. This approximation is consistent with the strong manifestation of this band within the shallow $\sim$1nm ARPES penetration depth \cite{WrayCuBiSePRB}, and with simulations that assume a sharp z-axis gradient for the surface dielectric constant \cite{WrayPN}. While there is strong evidence that dielectric permeability near the surface is renormalized after electropositive surface deposition, a more shallow gradient would cause the surface states to be less localized \cite{HofmannRashba2DEG}.

In summary, we have presented measurements of surface electronic structure as a function of Cu deposition on the electron doped topological insulator Cu$_x$Bi$_2$Se$_3$. These measurements are interpreted with the MTFA to present a broader discussion of how chemical and electrical gating of TI surface states may interact with surface electron density. We propose that carefully tuned surface bias potentials within the explored perturbation energy range could make it possible to enhance the proximity effects of bulk electronic ordering at the surface. The unusual charge inhomogeneity at the surface of topological insulators may also allow a method for creating tunable repulsive forces that could be useful to control the motion of superconducting vortices. For intrinsically homogeneous surfaces, the same types of external perturbation would likely suppress bulk-derived properties at the crystal surface and pin vortices in place.

\textbf{Acknowledgements:}

M.Z.H. is supported by U.S.DOE/BES grant number DE-FG02-05ER46200. The Advanced Light Source is supported by the Director, Office of Science, Office of Basic Energy Sciences, of the U.S. Department of Energy under Contract No. DE-AC02-05CH11231.

$\\$

\end{document}